\documentclass{aa}

\usepackage{upgreek}
\usepackage{txfonts}
\usepackage{graphicx}
\usepackage{xcolor}
\usepackage[colorlinks=true, linkcolor=blue, citecolor=blue, urlcolor=blue]{hyperref}
\usepackage{natbib}

\bibpunct{(}{)}{;}{a}{}{,} 

\newcommand{\txa}{{\text{a}}}
\newcommand{\txb}{{\text{b}}}
\newcommand{\txc}{{\text{c}}}
\newcommand{\txd}{{\text{d}}}
\newcommand{\txe}{{\text{e}}}
\newcommand{\calE}{{\cal{E}}}
\newcommand{\calN}{{\cal{N}}}
\newcommand{\calS}{{\cal{S}}}
\newcommand{\rb}{r_{\text{b}}}

\DeclareMathOperator{\erf}{erf}

\begin{document}

\title{{\tt{SpheCow}}: Flexible dynamical models for galaxies \\and dark matter haloes}

\author{Maarten Baes\inst{\ref{UGent}}, Peter Camps\inst{\ref{UGent}} \and Bert Vandenbroucke\inst{\ref{UGent}}}

\offprints{M.~Baes, \email{maarten.baes@ugent.be}}

\institute{
Sterrenkundig Observatorium, Universiteit Gent, Krijgslaan 281 S9, 9000 Gent, Belgium
\label{UGent}
}

\date{\today}

\abstract{Simple but flexible dynamical models are useful for many purposes, including serving as the starting point for more complex models or numerical simulations of galaxies, clusters, or dark matter haloes. We present {\tt{SpheCow}}, a new light-weight and flexible code that allows one to easily explore the structure and dynamics of any spherical model. Assuming an isotropic or Osipkov-Merritt anisotropic orbital structure, the code can automatically calculate the dynamical properties of any model with either an analytical density profile or an analytical surface density profile as starting point. We have extensively validated {\tt{SpheCow}} using a combination of comparisons to analytical and high-precision numerical calculations, as well as the calculation of inverse formulae. {\tt{SpheCow}} contains readily usable implementations for many standard models, including the Plummer, Hernquist, NFW, Einasto, S\'ersic, and Nuker models. The code is publicly available as a set of C++ routines and as a Python module, and it is designed to be easily extendable, in the sense that new models can be added in a straightforward way. We demonstrate this by adding two new families of models in which either the density slope or the surface density slope is described by an algebraic sigmoid function. We advocate the use of the {\tt{SpheCow}} code to investigate the full dynamical structure for models for which the distribution function cannot be expressed analytically and to explore a much wider range of models than is possible using analytical models alone.}


\maketitle

\section{Introduction}

In order to investigate the structure and evolution of self-gravitating systems such as galaxies, clusters, and dark matter haloes, a characterisation of their structure in full 6D phase space is of paramount importance. Thanks to the improving observational data, novel modelling techniques, and increasing computational power, we now have various advanced dynamical modelling tools at our disposal \citep[e.g.][]{2015ApJS..216...29B, 2019MNRAS.482.1525V, 2020ApJ...889...39V, 2021MNRAS.500.1437N}. Nevertheless, relatively simple spherically symmetric models remain useful and important. They can act as an environment in which different physical processes can be investigated, or against which new modelling or data analysis techniques can be explored. Moreover, they are often the starting point for the creation of more complex and realistic models or full-scale numerical simulations. 

Simple models have roughly become synonymous with analytical models, that is, models in which many properties can be calculated in terms of elementary or special functions. One approach to generate such simple models is by choosing a simple form for the distribution function. Once the distribution function is defined, the density profile $\rho(r)$ and potential $\Psi(r)$ of the self-consistent model can be determined by integrating the distribution function over velocity space and solving Poisson's equation. Famous models based on this approach include the singular isothermal sphere \citep{2008gady.book.....B}, the King model \citep{1966AJ.....71...64K}, and the Michie model \citep{1963MNRAS.125..127M, 1963MNRAS.126..269M}. Models with a simple analytical distribution function are very useful for elaborate calculations in which the precise form of the distribution function is not the major uncertainty, for example to compute the rate of microlensing events in dark matter haloes \citep{1995ApJ...449...28A} or to predict the signal in direct dark matter detection experiments \citep{2019PhRvD..99b3012E}. The disadvantage of this approach is that the density and potential are often not analytically tractable and/or do not necessarily provide a realistic representation for actual galaxies or dark matter haloes.

As an alternative to this approach, it is also common to set up dynamical models with the mass density profile $\rho(r)$ as a starting point. If the density is sufficiently simple, other quantities can often be expressed analytically as well, such as the cumulative mass profile and the gravitational potential. For a number of popular models, such as the NFW model \citep{1997ApJ...490..493N} or the Einasto model \citep{1965TrAlm...5...87E}, this is how far the analyticity goes \citep{2001MNRAS.321..155L, 2005MNRAS.358.1325C}. For a fairly limited set of models, analytical expressions are also available for the phase-space distribution function, usually under the assumption of a simple orbital structure such as velocity isotropy. Some of the most popular spherical models belong to this category, such as the Plummer model \citep{1911MNRAS..71..460P}, the isochrone sphere \citep{1959AnAp...22..126H}, the Jaffe model \citep{1983MNRAS.202..995J}, the Hernquist model \citep{1990ApJ...356..359H}, and the family of $\gamma$-models \citep{1993MNRAS.265..250D, 1994AJ....107..634T}. Finally, for a very restricted subset of these models more general anisotropic distribution functions and/or line profiles can be calculated analytically \citep[e.g.][]{1987MNRAS.224...13D, 2002A&A...393..485B, 2007A&A...471..419B}.

A disadvantage of models with a simple density profile as starting point is that their surface density profile on the plane of the sky\footnote{Throughout this paper, we assume a constant mass-to-light ratio, such that surface brightness and surface density are equivalent quantities.} is not necessarily equally simple \citep[e.g.][]{1996MNRAS.278..488Z, 2012A&A...540A..70R}. An interesting alternative is to consider models with a given surface density profile as a starting point. This approach has the advantage that specific functional forms can be chosen that closely reproduce the observed surface brightness distribution of galaxies and other systems. Examples of this type of models include the S\'ersic model \citep{1968adga.book.....S, 2005PASA...22..118G} and the Nuker model \citep{1995AJ....110.2622L}. The disadvantage is that the density needs to be calculated through an integral along the line of sight, and that this integration does not necessarily yield a simple expression. In particular, for the models listed above, the density profile can only be expressed in terms of very special functions \citep{2002A&A...383..384M, 2011A&A...525A.136B, 2011A&A...534A..69B, 2020A&A...634A.109B}. For other dynamical properties that typically require integrations of the density, finding explicit expressions is hopeless. 

While completely analytical models have their obvious benefits, it would be a pity to only focus on the limited set of models that belong to this category. We advocate the use of other models  as well, especially the much broader class of all models with an analytical density profile or surface density profile. Even when no analytical expressions are available for the most important dynamical properties, all of these properties can be calculated numerically in a relatively straightforward way.

In this paper we present a new light-weight and flexible code, {\tt{SpheCow}}, that sets up a spherical dynamical model with either an analytical density profile $\rho(r)$ or an analytical surface density profile $\Sigma(R)$ as starting point. For either of these two options, the {\tt{SpheCow}} code automatically calculates basic intrinsic and on-sky properties, as well as dynamical properties assuming either an isotropic or an Osipkov-Merritt type anisotropic orbital structure. This offers the possibility to easily calculate and investigate the full dynamical structure for models for which the distribution function cannot be expressed analytically, and to explore a much wider range of models than is possible using analytical models alone. In particular, such a light-weight tool can help to foster the use of more realistic models as starting points for more advanced modelling, at the expense of the simpler but often less realistic alternatives.

This paper is organised as follows. In Sect.~{\ref{DynProp.sec}} we discuss how the most important quantities, both intrinsic and on the plane of the sky, can be calculated when one has a closed expression for either the density (Sect.~{\ref{modsrho.sec}}) or the surface density (Sect.~{\ref{modssb.sec}}). This section reviews many well-known equations, but also presents some shortcuts that can be used to convert double to single integrals. In Sect.~{\ref{SpheCow.sec}} we present the {\tt{SpheCow}} code in which we have implemented these formulae. We present the numerical integration algorithm that is the core of the code, and we describe the design of the code. In Sect.~{\ref{Validation.sec}} we discuss the validation of the code. In Sect.~{\ref{Sigmoid.sec}} we present a simple application of the {\tt{SpheCow}} code. We introduce two new general families of models, in which the logarithmic slope of either the mass density or the surface density profile is characterised by a sigmoid function. We discuss and summarise our results in Sect.~{\ref{Summary.sec}}.

\section{The calculation of dynamical properties}
\label{DynProp.sec}

We assume that a spherically symmetric model is either characterised by an analytical density profile $\rho(r)$ or an analytical surface density profile $\Sigma(R)$. For either of these two cases, we discuss how the most important properties of the model can be calculated in an efficient way. In the remainder of this section, we assume that we have to our disposal a suitable numerical integration scheme that can efficiently calculate any integrals over semi-infinite or infinite intervals. This aspect will be discussed in Sect.~{\ref{NumIntegrations.sec}} and is considered as a given for the time being.

\subsection{Models defined by an analytical density profile}
\label{modsrho.sec}

\subsubsection{Basic properties}

In most cases, the starting point of a model is the 3D mass density $\rho(r)$. If an analytical form of $\rho(r)$ is known, we can also directly analytically calculate the first-order and second-order derivatives, and therefore also the negative logarithmic density slope 
\begin{equation}
\gamma(r) = -\frac{\txd \ln \rho}{\txd \ln r}(r) = -\frac{r\,\rho'(r)}{\rho(r)}.
\end{equation}
With $\rho(r)$ given, we can calculate the total mass
\begin{equation}
M = 4\pi \int_0^\infty \rho(u)\,u^2\,\txd u,
\label{Mtot}
\end{equation}
the cumulative mass distribution
\begin{equation}
M(r) = 4\pi \int_0^r \rho(u)\,u^2\,\txd u,
\label{M(r)}
\end{equation}
and the positive gravitational potential 
\begin{equation}
\Psi(r) 
= 
\frac{GM(r)}{r} + 4\pi\,G\int_r^\infty \rho(u)\,u\,\txd u.
\label{Psi(r)}
\end{equation}
The circular velocity curve is
\begin{equation}
v_{\text{c}}(r) = \frac{G M(r)}{r}.
\end{equation}
The total potential energy is calculated as
\citep{2008gady.book.....B, 2019A&A...630A.113B}
\begin{equation}
W_{\text{tot}} = -4\pi G \int_0^\infty \rho(u)\,M(u)\,u\,\txd u.
\end{equation}
The surface density profile can be calculated using an integration along the line of sight,
\begin{equation}
\Sigma(R) = 2\int_R^\infty \frac{\rho(u)\,u\,\txd u}{\sqrt{u^2-R^2}}.
\label{rhotoI}
\end{equation}
In a similar way as for the density profile, it is often interesting to characterise the negative logarithmic slope of the surface density profile \citep[e.g.][]{1995AJ....110.2622L, 2007ApJ...664..226L},
\begin{equation}
\gamma_{\text{p}}(R) = -\frac{\txd \ln \Sigma}{\txd \ln R}(R) = -\frac{R\,\Sigma'(R)}{\Sigma(R)}.
\end{equation}
This expression requires the derivative of the surface density profile, for which we have, in general, no closed analytical expression. One way to automatically calculate it for an arbitrary model with a given density profile $\rho(r)$ is to directly apply numerical differentiation to the surface density profile calculated numerically through expression~(\ref{rhotoI}). A more convenient and accurate way, however, is to directly take the formal derivative of expression~(\ref{rhotoI}). Using the substitution $u=R\sec\theta$ we have
\begin{equation}
\Sigma(R) = 2\int_0^{\pi/2} \rho(R\sec\theta)\,R\sec^2\theta\,\txd\theta.
\end{equation}
Taking the derivative with respect to $R$ and transforming it back to $u$, we find
\begin{equation}
\Sigma'(R) = 2\int_R^\infty \frac{[\rho(u)+ u\,\rho'(u)]\,u\,\txd u}{R \sqrt{u^2-R^2}},
\label{Sigma'}
\end{equation}
where $\rho'(r)$ is the first-order derivative of the density. The cumulative surface mass profile is the mass within a circular surface density contour on the sky with radius $R$,
\begin{equation}
M_{\text{p}}(R) = 2\pi\int_0^R \Sigma(u)\,u\,\txd u.
\label{Mp-gen}
\end{equation}
Inserting expression~(\ref{rhotoI}) and changing the order of the integration, we get \citep{2010A&A...520A..30M}
\begin{equation}
M_{\text{p}}(R) = M - 4\pi\int_R^\infty \rho(u)\,u\sqrt{u^2-R^2}\,\txd u.
\label{Mp-res}
\end{equation}
This expression involves a combination of two single integrals, one of which only needs to be calculated once, rather than a double integration.

\subsubsection{Dynamical properties for an isotropic orbital structure}

It is well-known that each spherical potential-density pair corresponds to a unique dynamical model with an isotropic velocity distribution \citep{1986PhR...133..217D, 2008gady.book.....B}. The velocity dispersion profile $\sigma_{\text{iso}}^2(r)$ of this model can be found through the solution of the Jeans equation,
\begin{equation}
\rho(r)\,\sigma_{\text{iso}}^2(r) = G\int_r^\infty \frac{\rho(u)\,M(u)\,\txd u}{u^2}.
\label{sigma2}
\end{equation}
The total kinetic energy is calculated as 
\begin{equation}
K_{\text{tot}} = 6\pi\int_0^\infty \rho(u)\,\sigma_{\text{iso}}^2(u)\,u^2\,\txd u.
\end{equation}
The line-of-sight velocity or projected dispersion $\sigma_{{\text{p}},{\text{iso}}}^2(R)$ is found as
\begin{equation}
\Sigma(R)\,\sigma_{{\text{p}},{\text{iso}}}^2(R) 
= 
2 \int_R^\infty \frac{\rho(u)\,\sigma_{\text{iso}}^2(u)\,u\,\txd u}{\sqrt{u^2-R^2}}.
\end{equation}
We can introduce expression~(\ref{sigma2}) into this equation and change the order of integration \citep{1997A&A...321..111P}. This yields the equivalent expression
\begin{equation}
\Sigma(R)\,\sigma_{{\text{p}},{\text{iso}}}^2(R) 
=
2G \int_R^\infty \frac{\rho(u)\,M(u) \sqrt{u^2-R^2}\,\txd u}{u^2}.
\label{sigmap_iso}
\end{equation}
The isotropic distribution function, which only depends on the (positive) binding energy per unit mass $\calE$, can be found through Eddington's formula \citep{1916MNRAS..76..572E},
\begin{equation}
f_{\text{iso}}(\calE) = \frac{1}{2\!\sqrt2\,\pi^2} \int_0^\calE \frac{\tilde\rho''(\Psi)\,\txd\Psi}{\sqrt{\calE-\Psi}},
\label{Eddington}
\end{equation}
with $\tilde\rho(\Psi)$ the augmented density, that is\ the density written as a function of the potential. Following \citet{1982MNRAS.200..951B}, we transform this integral via the substitution $\Psi=\Psi(u)$, where we use the identity
\begin{equation}
\frac{\txd\Psi(r)}{\txd r} = -\frac{G M(r)}{r^2}.
\end{equation}
This results in 
\begin{subequations}
\begin{equation}
f_{\text{iso}}(\calE) = \frac{1}{2\!\sqrt2\,\pi^2} \int_{r_\calE}^\infty \frac{\Delta(u)\,\txd u}{\sqrt{\calE-\Psi(u)}},
\label{fEDelta}
\end{equation}
with 
\begin{equation}
\Delta(r) = \frac{r^2}{GM(r)}\left[\rho''(r) + \rho'(r)
\left(\frac{2}{r} - \frac{4\pi\,\rho(r)\,r^2}{M(r)}\right)
\right],
\label{defDelta}
\end{equation}
\end{subequations}
and $r_\calE$ implicitly defined by the equation $\Psi(r_\calE) = \calE$. 

The differential energy distribution represents the distribution of mass as a function of the binding energy $\calE$. For isotropic systems, the differential energy distribution can be written as
\begin{equation}
\calN_{\text{iso}}(\calE) = f_{\text{iso}}(\calE)\,g_{\text{iso}}(\calE),
\end{equation}
with $g_{\text{iso}}(\calE)$ the density-of-states function, defined as
\begin{equation}
g_{\text{iso}}(\calE) 
= 
-16\!\sqrt2\,\pi^2 \int_\calE^{\Psi_0} r^2\,\frac{\txd r}{\txd\Psi} \sqrt{\Psi-\calE}\,\txd\Psi
\label{dos}
\end{equation}
with $r = r(\Psi)$ the radius written as a function of the binding potential, and $\Psi_0$ the depth of the potential well. This equation can be recast in the form 
\begin{equation}
g_{\text{iso}}(\calE) 
=
16\!\sqrt2\,\pi^2 \int_0^{r_\calE} u^2 \sqrt{\Psi(u)-\calE}\,\txd u.
\end{equation}

\subsubsection{Dynamical properties for an Osipkov-Merritt orbital structure}

An isotropic orbital structure is the simplest option and it is interesting from a mathematical point of view. However, galaxies and dark matter haloes are not necessarily characterised by this simple orbital structure. In fact, most numerical N-body simulations predict that dark matter haloes are roughly isotropic in the central regions, but radially anisotropic in the outer regions \citep{2001ApJ...563..483T, 2004MNRAS.352..535D, 2011MNRAS.415.3895L, 2012ApJ...752..141L, 2013MNRAS.434.1576W, 2016MNRAS.462..663B}. 

\citet{1979PAZh....5...77O} and \citet{1985AJ.....90.1027M} independently discovered that dynamical systems with a spheroidal velocity distribution have the interesting property that their distribution function can be calculated in a similar way as models with an isotropic distribution function. The now so-called Osipkov-Merritt models are characterised by an orbital structure that is isotropic in the centre and completely radially anisotropic at large radii. More concretely, the anisotropy profile $\beta(r)$ is
\begin{equation}
\beta(r) \equiv 1-\frac{\sigma^2_\theta(r)}{\sigma_r^2(r)} = \frac{r^2}{r^2+r_\txa^2},
\end{equation}
with $r_\txa$ a freely chosen anisotropy radius. In this case, the velocity dispersions are obviously not the same in the three principle directions. For a given density profile and anisotropy radius, the radial velocity dispersion profile can be calculated as \citep{2005MNRAS.363..705M}
\begin{equation}
\rho(r)\,\sigma_{r,{\text{om}}}^2(r) 
=
\frac{G r_\txa^2}{r^2+r_\txa^2} \int_r^\infty \frac{\rho_Q(u)\,M(u)\,\txd u}{u^2},
\label{sigmar2Q}
\end{equation}
with $\rho_Q(r)$ defined as
\begin{equation}
\rho_Q(r) = \left(1+ \frac{r^2}{r_\txa^2}\right) \rho(r).
\end{equation}
The velocity dispersions in the two tangential directions are given by
\begin{equation}
\sigma_{\theta,{\text{om}}}^2(r) = \sigma_{\phi,{\text{om}}}^2(r) 
= \frac{r_\txa^2}{r^2+r_\txa^2}\,\sigma_{r,{\text{om}}}^2(r).
\end{equation}
In this case, the total kinetic energy is calculated as 
\begin{align}
K_{\text{tot}} 
&= 
2\pi\int_0^\infty \rho(u) \left[\sigma_{r,{\text{om}}}^2(u) + \sigma_{\theta,{\text{om}}}^2(u) + \sigma_{\phi,{\text{om}}}^2(u) \right] u^2\,\txd u
\nonumber \\
&=
2\pi\int_0^\infty \left(\frac{u^2+3 r_\txa^2}{u^2+r_\txa^2}\right)\,\rho(u)\,\sigma_{r,{\text{om}}}^2(u)\,u^2\,\txd u.
\end{align}
The general expression for the projected velocity dispersion for an anisotropic spherical model is \citep{1982MNRAS.200..361B, 1987MNRAS.224...13D}
\begin{equation}
\Sigma(R)\,\sigma_{\text{p}}^2(R) 
= 
2 \int_R^\infty \left[1-\frac{R^2}{u^2}\,\beta(u)\right]\frac{\rho(u)\,\sigma_r^2(u)\,u\,\txd u}{\sqrt{u^2-R^2}},
\end{equation}
which in the case of an Osipkov-Merritt orbital structure reduces to
\begin{equation}
\Sigma(R)\,\sigma_{{\text{p}},{\text{om}}}^2(R) 
= 
2 \int_R^\infty 
\left(\frac{u^2+r_\txa^2-R^2}{u^2+r_\txa^2}\right) 
\frac{\rho(u)\,\sigma_{r,{\text{om}}}^2(u)\,u\,\txd u}{\sqrt{u^2-R^2}}.
\end{equation}
In principle, this expression can directly be implemented, but the integrand depends on $\sigma_{r,{\text{om}}}^2(u)$, implying that we get at least a double integral (and a triple integral if the mass profile needs to be calculated numerically). A more convenient form can be obtained by substituting Eq.~(\ref{sigmar2Q}) and changing the order of integration in the resulting expression \citep{2005MNRAS.363..705M, 2014MNRAS.442.3284A}. This results in 
\begin{subequations}
\label{sigmap_om}
\begin{equation}
\Sigma(R)\,\sigma_{{\text{p}},{\text{om}}}^2(R) 
= 
G\int_R^\infty \frac{w(u,R)\,\rho(u)\,M(u)\,\txd u}{u^2},
\end{equation}
with
\begin{multline}
w(u,R) = \left(\frac{u^2+r_\txa^2}{R^2+r_\txa^2}\right)
\\
\times
\left( 
\frac{R^2+2r_\txa^2}{\sqrt{R^2+r_\txa^2}}
\arctan\!\sqrt{\frac{u^2-R^2}{R^2+r_\txa^2}}
- \frac{R^2\sqrt{u^2-R^2}}{u^2+r_\txa^2} 
\right).
\end{multline}
\end{subequations}
The distribution function of anisotropic dynamical models can generally written as $f(\calE,L)$ with $L$ the total angular momentum per unit mass. For the special case of an Osipkov-Merritt orbital structure, the distribution function depends on $\calE$ and $L$ only through the combination
\begin{equation}
Q = \calE - \frac{L^2}{2r_\txa^2},
\end{equation}
that is, $f_{\text{om}}(\calE,L) = f_{\text{om}}(Q)$, with the additional constraint that $f_{\text{om}}(Q) = 0$ for $Q\leqslant 0$. For a given density profile, the function $f_{\text{om}}(Q)$ can be determined using a formula very similar to the Eddington relation \citep{1979PAZh....5...77O, 1985AJ.....90.1027M},
\begin{equation}
f_{\text{om}}(Q) = \frac{1}{2\!\sqrt2\,\pi^2} \int_0^Q \frac{\tilde\rho''_Q(\Psi)\,\txd\Psi}{\sqrt{Q-\Psi}}.
\end{equation}
As for the isotropic case, this expression can be recast in a form that is more suitable for numerical integration,
\begin{subequations}
\begin{equation}
f_{\text{om}}(Q) = \frac{1}{2\!\sqrt2\,\pi^2} \int_{r_Q}^\infty \frac{\Delta_Q(u)\,\txd u}{\sqrt{Q-\Psi(u)}},
\label{fQDeltaQ}
\end{equation}
with $r_Q$ determined implicitly by $\Psi(r_Q)=Q$, and 
\begin{equation}
\Delta_Q(r) = \frac{r^2}{GM(r)}\left[\rho_Q''(r) + \rho_Q'(r)
\left(\frac{2}{r} - \frac{4\pi\,\rho_Q(r)\,r^2}{M(r)}\right)
\right].
\label{defDeltaQ}
\end{equation}
\end{subequations}
A general calculation of the differential energy distribution for Osipkov-Merritt orbital structure is non-trivial. However, one can define a probability density function $\calN_{\text{om}}(Q)$ for the mass as a function of $Q$, which we refer to as the pseudo-differential energy distribution. \citet{1991MNRAS.253..414C} demonstrated that it can be found as 
\begin{equation}
\calN_{\text{om}}(Q) = f_{\text{om}}(Q)\,g_{\text{om}}(Q),
\end{equation}
with $g_{\text{om}}(Q)$ a pseudo-density-of-states function, 
\begin{equation}
g_{\text{om}}(Q) 
=
16\!\sqrt2\,\pi^2 \int_0^{r_Q} u^2\left(1+\frac{u^2}{r_\txa^2}\right)^{-1} \sqrt{\Psi(u)-Q}\,\txd u.
\end{equation}

\subsection{Models defined by an analytical surface density profile}
\label{modssb.sec}

As discussed in the Introduction, another class of models starts from an observed surface density profile $\Sigma(R)$, with $R$ the radius on the plane of the sky. As can be seen in the previous section, the fundamental quantity that appeared in nearly all the formulae is the density $\rho(r)$. The connection between density and surface density is given by equation~(\ref{rhotoI}), which can be inverted using the standard Abel inversion formula \citep{2008gady.book.....B, 2010MNRAS.401.2433M}
\begin{equation}
\rho(r) = -\frac{1}{\pi} \int_r^\infty \frac{\Sigma'(u)\,\txd u}{\sqrt{u^2-r^2}}.
\label{Itorho}
\end{equation}
In this formula, $\Sigma'(R)$ is the derivative of the surface density profile, which is also known analytically since we assumed that $\Sigma(R)$ is an analytical function. For a given analytical surface density profile, we can numerically integrate expression~(\ref{Itorho}) to obtain $\rho(r)$, and subsequently use this expression in all other expression. However, we can avoid the use of numerical double integrals for the calculation of the mass profile and gravitational potential by inserting the expression~(\ref{Itorho}) into the general Eqs.~(\ref{M(r)}) and (\ref{Psi(r)}), and changing the order of integrations. After some manipulation we find for the mass profile
\begin{subequations}
\label{Mfromsb}
\begin{equation}
M(r)
=
-\pi\left[\int_0^r \Sigma'(u)\,u^2\, \txd u 
+ \int_r^\infty \Sigma'(u)\,w_-(u,r)\, \txd u \right],
\end{equation}
where we have set
\begin{equation}
w_-(u,r) = \frac{2}{\pi}\left[u^2\arctan\left(\frac{r}{\sqrt{u^2-r^2}}\right)-r\sqrt{u^2-r^2}\right].
\end{equation}
\end{subequations}
For the gravitational potential we find in a similar way
\begin{subequations}
\label{Psifromsb}
\begin{equation}
\Psi(r) 
= 
-\frac{\pi\,G}{r} 
\left[
\int_0^r \Sigma'(u)\,u^2\, \txd u 
+ \int_r^\infty \Sigma'(u)\,w_+(u,r)\,\txd u \right],
\end{equation}
with now
\begin{equation}
w_+(u,r) = \frac{2}{\pi}\left[u^2\arctan\left(\frac{r}{\sqrt{u^2-r^2}}\right)+r\sqrt{u^2-r^2}\right].
\end{equation}
\end{subequations}
For all other properties, corresponding to either an isotropic or an Osipkov-Merritt orbital distribution, we directly use the expressions from the previous subsection, in which we substitute the (numerically determined) expressions (\ref{Itorho}), (\ref{Mfromsb}), and (\ref{Psifromsb}) for density, mass profile, and potential. The only additional quantities that we need are the first and second order derivatives of the density, which are necessary to calculate the slope of the density profile and the distribution function. Rather than applying numerical differentiation to the numerically determined density profile (\ref{Itorho}), it is more straightforward and far more accurate to use the expressions
\begin{gather}
\rho'(r) = -\frac{1}{\pi} \int_r^\infty \frac{\Sigma''(u)\,u\,\txd u}{r\sqrt{u^2-r^2}},
\\
\rho''(r) = -\frac{1}{\pi} \int_r^\infty \frac{\Sigma'''(u)\,u^2\,\txd u}{r^2\sqrt{u^2-r^2}},
\end{gather}
with $\Sigma''(R)$ and $\Sigma'''(R)$ the second and third order derivatives of the surface density profile, for which we assume exact expressions are available. These formulae can be obtained from Eq.~(\ref{Itorho}) in a similar way as we derived Eq.~(\ref{Sigma'}) from Eq.~(\ref{rhotoI}).

\section{The {\tt{SpheCow}} code}
\label{SpheCow.sec}

\begin{table*}
\caption{Models currently implemented in the {\tt{SpheCow}} code. Models defined by an analytical density profile are subclasses of the general {\tt{DensityModel}} class, models defined by an analytical surface density profile are subclasses of the general {\tt{SurfaceDensityModel}} class.} 
\label{models.tab}
\centering
\begin{tabular}{lll}
\hline\hline\\
name & {\tt{SpheCow}} class & references \\[1em]
\hline \\
Models defined by $\rho(r)$ & {\tt{DensityModel}} \\[0.5em]
\qquad Broken power-law & {\tt{BPLModel}} & \citet{2020ApJ...892...62D, 2021MNRAS.503.2955B} \\
\qquad Burkert & {\tt{BurkertModel}} & \citet{1995ApJ...447L..25B, 2000ApJ...537L...9S} \\
\qquad Einasto & {\tt{EinastoModel}} & \citet{1965TrAlm...5...87E, 2005MNRAS.358.1325C} \\
\qquad $\gamma$--model & {\tt{GammaModel}} & \citet{1993MNRAS.265..250D, 1994AJ....107..634T}; \\
& & \quad{\citet{1995MNRAS.276.1131C, 2004MNRAS.351...18B}} \\
\qquad Hernquist & {\tt{HernquistModel}} & {\citet{1990ApJ...356..359H, 2002A&A...393..485B}} \\
\qquad Hypervirial & {\tt{HypervirialModel}} & {\citet{1979AZh....56..976V, 2005MNRAS.360..492E}} \\
\qquad Isochrone & {\tt{IsochroneModel}} & \citet{1959AnAp...22..126H, 1960AnAp...23..474H, 1986MNRAS.221P..13M} \\
\qquad Jaffe & {\tt{JaffeModel}} & \citet{1983MNRAS.202..995J, 1985MNRAS.214P..25M} \\ 
\qquad Navarro, Frenk \& White & {\tt{NFWModel}} & \citet{1997ApJ...490..493N, 2001MNRAS.321..155L} \\
\qquad Perfect sphere & {\tt{PerfectSphereModel}} & \citet{1985MNRAS.216..273D} \\
\qquad Plummer & {\tt{PlummerModel}} & \citet{1911MNRAS..71..460P, 1987MNRAS.224...13D}; \\
& & \quad{\citet{1985AJ.....90.1027M, 1991MNRAS.253..414C}} \\
\qquad Zhao & {\tt{ZhaoModel}} & \citet{1996MNRAS.278..488Z}  \\[0.5em]
Models defined by $\Sigma(R)$ & {\tt{SurfaceDensityModel}} \\[0.5em]
\qquad de Vaucouleurs & {\tt{DeVaucouleursModel}} & {\citet{1948AnAp...11..247D, 1982MNRAS.200..951B}} \\
\qquad Nuker & {\tt{NukerModel}} & {\citet{1995AJ....110.2622L, 2020A&A...634A.109B}} \\
\qquad S\'ersic & {\tt{SersicModel}} & {\citet{1968adga.book.....S, 2005PASA...22..118G}}; \\
& & \quad{\citet{1991A&A...249...99C, 2019A&A...626A.110B}} \\[1.5em]
\hline\hline
\end{tabular}
\end{table*}

We have designed a C++ software module called {\tt{SpheCow}} to automatically calculate the dynamical properties as described above for any spherical model. Models can be defined either by their density profile, or by their surface density profile. In both cases, the code can be used to calculate the basic model properties and the most important dynamical properties, for both isotropic and Osipkov-Merritt-type anisotropic orbital structures. 

\subsection{Numerical integrations}
\label{NumIntegrations.sec}

At the core of the {\tt{SpheCow}} code is a generic numerical integration scheme that is appropriate for all models, whether they are based on density or surface density. As can be seen from the set of formulae in the previous section, we essentially encounter only two types of integrals: all integrations run over the variable $u$, which can represent either the spherical radius or the projected radius on the plane of the sky, and they either run on an interval $[0,r]$ or an interval $[r,\infty[$. So in principle we only have to implement a strategy to numerically calculate such integrals. A more refined strategy is required, however, to ensure that we get accurate results for a wide variety of galaxy models. Realistic models are typically characterised by a particular scale radius $\rb$, which often corresponds to the division between an inner and an outer profile. It therefore seems logical to split any integration over an interval that contains $\rb$ into two separate integrals. This is particularly necessary for models in which there is a (relatively) sharp break in the density profile at $r=\rb$.

This implies that we need to consider four different types of integrals, namely integrals on the intervals $[0,r]$ or $[r,\rb]$ if $r<\rb$, or on the intervals $[\rb,r]$ or $[r,\infty[ $ if $r>\rb$. For each of these integrals we use a substitution to turn it into an equivalent integration that is more suitable for a Gauss-Legendre quadrature scheme. In the case $r<\rb$, we use
\begin{gather}
\int_0^r X(u)\,\txd u = r \int_0^{\pi/2} X(r\sin\theta) \cos\theta\,\txd\theta,
\label{int0r}
\\
\int_r^{\rb} X(u)\,\txd u = r \int_{\arcsin (r/\rb)}^{\pi/2} X(r\csc\theta) \cos\theta \csc^2\theta\,\txd\theta.
\end{gather}
For $r>\rb$ we use similar transformations,
\begin{gather}
\int_{\rb}^r X(u)\,\txd u = r \int_{\arcsin (\rb/r)}^{\pi/2} X(r\sin\theta) \cos\theta\,\txd\theta, \\
\int_r^\infty X(u)\,\txd u = r \int_0^{\pi/2} X(r\csc\theta) \cos\theta \csc^2\theta\,\txd\theta.
\label{intrinf}
\end{gather}
For integrals covering the entire intervals $[0,\rb]$ and $[\rb,\infty[$, we use the transformations~(\ref{int0r}) and (\ref{intrinf}), respectively, with $r\to\rb$,
\begin{gather}
\int_0^{\rb} X(u)\,\txd u = \rb \int_0^{\pi/2} X(\rb\sin\theta) \cos\theta\,\txd\theta,
\\
\int_{\rb}^\infty X(u)\,\txd u = \rb \int_0^{\pi/2} X(\rb\csc\theta) \cos\theta \csc^2\theta\,\txd\theta.
\end{gather}
To perform the actual integrations, we use a Gauss-Legendre quadrature formula. The integration points and the corresponding weights are pre-calculated, which makes the integrations fast and efficient. We use 128 integration points as our default, but this value can be chosen by the user. In general, higher orders correspond to higher accuracy but to an increased calculation time as well (see Sec.~{\ref{Accuracy.sec}} for a discussion).

\subsection{Code design}

For the design of the {\tt{SpheCow}} code we have deliberately chosen a modular and light-weight setup. The code essentially consists of a simple driver function, a {\tt{GaussLegendre}} class that takes care of the numerical integrations, and a general abstract {\tt{Model}} class that acts as the base class for all different models. This last class has two abstract subclasses, {\tt{DensityModel}} and {\tt{SurfaceDensityModel}}. The former is the base class for all different models defined by an analytical density profile, the latter for all models with an analytical surface density profile as starting point. 

Once the density or surface density and their derivatives are known, the calculation of all the photometric and dynamical properties is calculated via a direct implementation of the formulae as written in the previous Section. The only aspect where some special care is needed is the integrations that contain a factor $\Psi(r)-\Psi(u)$ in the denominator in the integrand. When $u-r=\epsilon$ is very small, this difference can effectively become zero and this can cause infinities. In this case we use the first two terms in the Taylor expansion,
\begin{equation}
\Psi(r)-\Psi(r+\epsilon) \approx \frac{GM(r)}{r^2}\,\epsilon + \left[2\pi\,\rho(r)-\frac{GM(r)}{r^3}\right] \epsilon^2.
\end{equation}

We have currently implemented a suite of models, listed in Table~{\ref{models.tab}}. There are many other models in the literature that we have not (yet) included in the code, including the generalised isochrone sphere \citep{2006AJ....131..782A}, the core-NFW model \citep{2012ApJ...759L..42P}, the doubloon models \citep{2014MNRAS.443..791E, 2015MNRAS.450..846E}, or the generalised Einasto model \citep{2020MNRAS.499.2426F}. However, the {\tt{SpheCow}} code is modular in design, and new models can easily be added, as separate subclasses that inherit from either the {\tt{DensityModel}} and {\tt{SurfaceDensityModel}} base classes. To add a model with an analytical density profile, the user needs to provide an implementation of the density, and its first and second order derivatives. All other quantities can be calculated numerically using the formulae and numerical integration schemes discussed above. However, in case closed analytical expressions are available, in particular for the mass profile or the potential, these can also be provided. This will help the calculation of other properties such as the distribution function, both in speed and accuracy. Similarly, to add a new model with an analytical surface density profile, a new subclass of the {\tt{SurfaceDensityModel}} class must be generated, and an implementation of the surface density profile $\Sigma(R)$, and its first, second, and third order derivatives are required. All other quantities can be calculated automatically.

\subsection{Availability}

{\tt{SpheCow}} is publicly available on GitHub (\url{https://github.com/mbaes/SpheCow}) as a set of C++ files. A Makefile and a readme file are available to help users with the installation and running of the code. We have also foreseen Python bindings and a Python script to convert the {\tt{SpheCow}} code to a Python module that can directly be used in Python code and notebooks. An Jupyter Notebook example is included to help users to get started. 

\section{Validation}
\label{Validation.sec}

\subsection{Accuracy of the numerical integrations}
\label{Accuracy.sec}

We have extensively validated the {\tt{SpheCow}} code. A first important aspect is a characterisation of the accuracy of the calculations, which we have done by comparing the results of the code against analytical results for a number of quantities and models where such analytical expressions exist. Very useful models for which analytical expressions exist for nearly all relevant quantities, both for an isotropic and an Osipkov-Merritt orbital structure, are the Plummer model, the Hernquist model, and the $\gamma$--model with $\gamma=\tfrac52$. For most other models implemented, at least some properties can be calculated analytically: analytical formulae can be found in the articles referenced in the last column of Table~{\ref{models.tab}}. We have compared the numerical results against the analytical ones where possible, and generally found excellent agreement. 
 
\begin{figure*}
\includegraphics[width=0.96\textwidth]{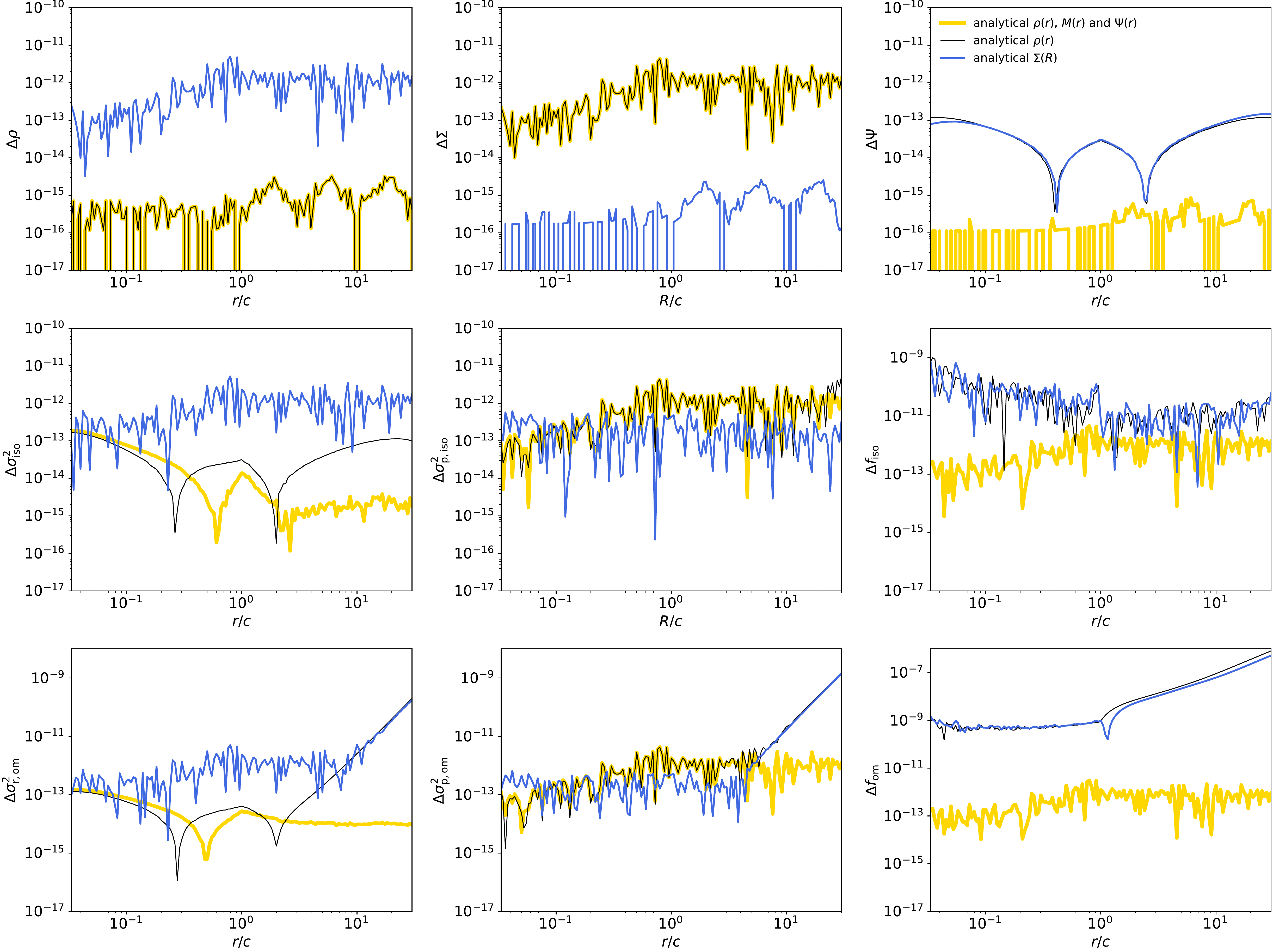}%
\caption{Absolute value of the relative error of the density, surface density, potential, velocity dispersion, projected dispersion, and distribution function of the Plummer model ($c$ represents the scale radius of the Plummer model). See text in Sec.~{\ref{Accuracy.sec}} for the meaning of the different curves in each panel.}
\label{Validation-Plummer.fig}
\end{figure*}

In Fig.~{\ref{Validation-Plummer.fig}} we illustrate the accuracy of the {\tt{SpheCow}} code by means of the Plummer model. In the nine panels we show the absolute value of the relative errors between the {\tt{SpheCow}} calculations and extended precision calculations based on the analytical formulae from \citet{1985AJ.....90.1027M} and \citet{1987MNRAS.224...13D}. The top row shows three basic properties: the density, the surface density, and the potential. The middle row shows the intrinsic dispersion, the projected dispersion, and the distribution function corresponding to an isotropic orbital structure, whereas the bottom row shows the same properties, but now for an Osipkov-Merritt orbital structure with $r_\txa=c$, with $c$ the scale radius of the Plummer model.

The thick yellow curves correspond to the Plummer model as implemented in the {\tt{SpheCow}} code. It is a subclass of the {\tt{DensityModel}} class, that is, it is defined by means of its analytical density profile. Since the Plummer model also has simple analytical expressions for the mass profile and the gravitational potential, these are also implemented in the {\tt{PlummerModel}} class, and this explains why the density and the potential have relative errors of the order of machine precision. The surface density profile, the intrinsic and projected dispersion profiles, and the distribution functions all involve just a single numerical integration. For the surface density and projected dispersion, the relative error is of the order of $10^{-12}$; for the intrinsic dispersion and the distribution functions, the relative error is even smaller.  

The thin black curves in Fig.~{\ref{Validation-Plummer.fig}} also represent {\tt{SpheCow}} calculations for a Plummer model, but now we have not used the specific Plummer implementations for the potential and mass profile. Both are now calculated numerically using our standard Gauss-Legendre integration scheme based on Eqs.~(\ref{M(r)}) and (\ref{Psi(r)}), and this results in a relative error of the order of $10^{-13}$. This does not affect the surface density profile and it only marginally increases the relative error of the dispersion profiles for the isotropic model. It does affect the accuracy of the dispersion profiles of the models with an Osipkov-Merritt orbital structure, especially at the outer, radially anisotropic regions. It also affects the relative error of the isotropic and Osipkov-Merritt distribution functions. For the isotropic case, the distribution function is now characterised by a relative error of the order of $10^{-10}$, especially at small radii. For the anisotropic distribution function, we see a larger loss of accuracy. In the inner, isotropic regions we have a relative error of about $10^{-9}$, in the outer radially anisotropic region, the relative error increases up to $10^{-6}$.

Finally, we have also implemented an alternative version of the Plummer model in {\tt{SpheCow}}, now as subclass of the {\tt{SurfaceDensityModel}} class. Indeed, the Plummer model is quite unique in that it also has a very simple analytical surface density profile. The results of this alternative model are shown as the blue lines in Fig.~{\ref{Validation-Plummer.fig}}. In this case, the density profile, mass profile, and potential all involve a single numerical integration, as shown in Eqs.~(\ref{Itorho}), (\ref{Mfromsb}), and (\ref{Psifromsb}). For the density, this results in a relative error of the order of $10^{-12}$, for the potential the accuracy is an order of magnitude better. Both the intrinsic and the line-of-sight dispersion profiles now require a double numerical integration, but still the relative error remains at the order of $10^{-12}$ for the intrinsic dispersion and even around $10^{-13}$ for the projected dispersion (except for the Osipkov-Merritt case at large radii). The relative error on the distribution functions does not alter significantly.

\begin{figure}
\includegraphics[width=0.45\textwidth]{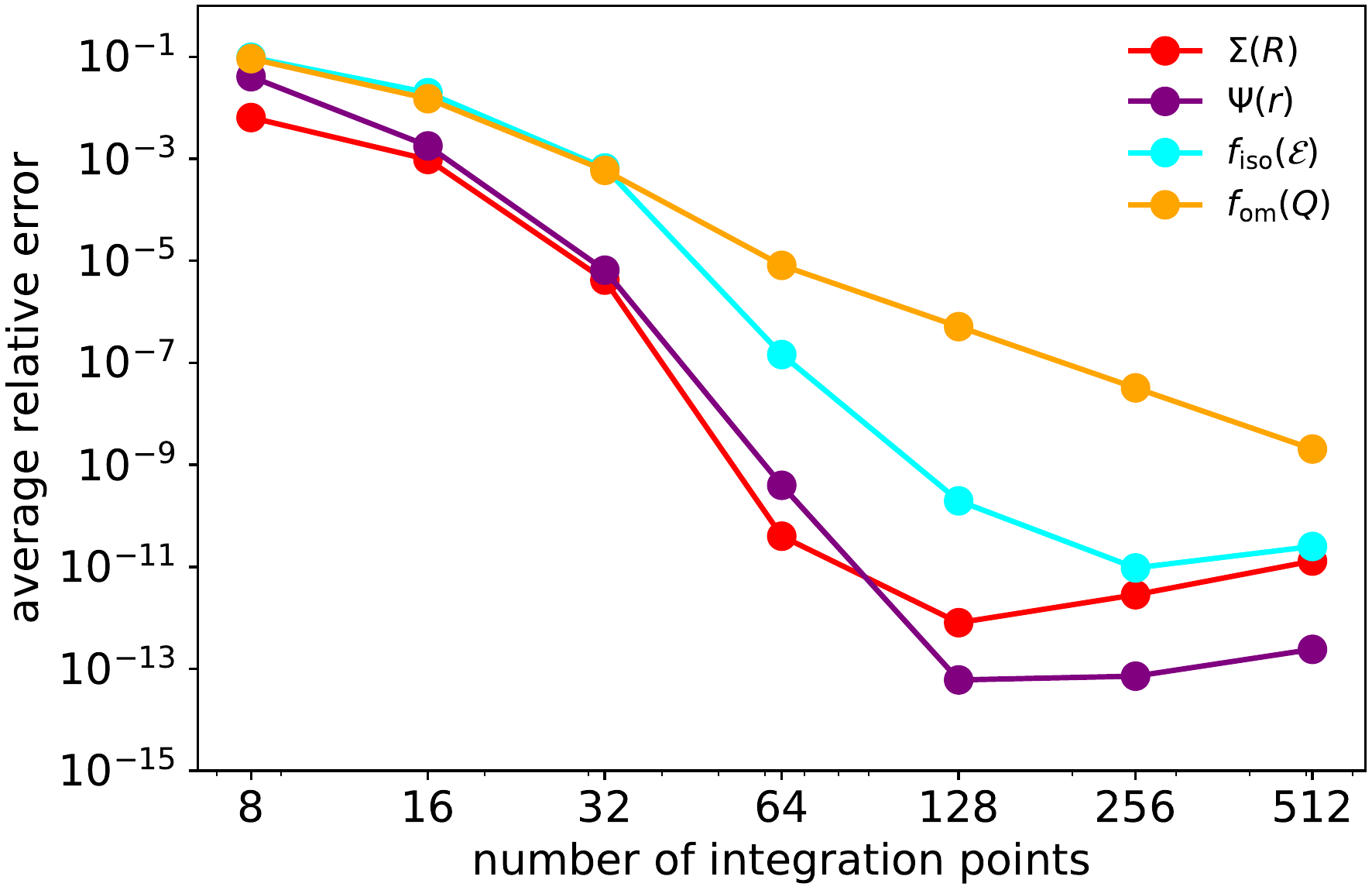}%
\caption{Average relative error on the calculation of the surface density, potential, and distribution function for the Plummer model as a function of the number of integration points in the Gauss-Legendre integration scheme.}
\label{IntegrationPoints.fig}
\end{figure}

For Fig.~{\ref{Validation-Plummer.fig}} we used our default Gauss-Legendre integration scheme with 128 nodes, but this number can be chosen freely by the user, as discussed in Sec.~{\ref{NumIntegrations.sec}}. In Fig.~{\ref{IntegrationPoints.fig}} we show how the relative error on the surface density, potential, and distribution function changes as a function of the number of integration points in the numerical integration scheme. To get a single representative relative error, we calculated the relative error corresponding to 201 points spread logarithmically between $r/r_{\text{b}}=10^{-2}$ and $10^2$, and we took the average value of the absolute values. We used a Plummer model without an explicit implementation of the potential and the mass profile for the calculations, that is the model corresponding to the thin black lines in Fig.~{\ref{Validation-Plummer.fig}}. The number of integration points is varied from 8 to 512. 

For the surface density and the potential, quantities that require just one and two individual integrations respectively, the accuracy of the {\tt{SpheCow}} calculation increases spectacularly as the number of integration points increases. For example, increasing this number from 32 to 64, the relative error decreases by roughly five orders of magnitude. As soon as the number of points reaches about 128, the accuracy does not improve any further. The calculation of the distribution function, both for an isotropic and an Osipkov-Merritt orbital structure, is more complex and involves nested numerical integrations. The accuracy of the calculations still increases strongly with increasing number of integration points, but not as outspoken as for the surface density and potential. The accuracy also keeps increasing beyond 128 integration points. 

In general, Fig.~{\ref{IntegrationPoints.fig}} and additional tests suggest that a Gauss-Legendre integration scheme with 128 integration points represents a good compromise between high accuracy and reasonable computation time (for quantities that involve a double integral, such as the distribution function, the computation time scales quadratically with the number of integration points). In case a reduced accuracy is sufficient, a slightly lower order can be considered. 

 \subsection{Consistency checks}
 
The {\tt{SpheCow}} code already contains a significant number of models and is designed to be extended with additional models in a straightforward way. Each new model requires the implementation of either the density and its first two derivatives, or the surface density and it first three derivatives. In order to check the consistency of the implementation of these properties, we can use the inverse formulae for many of the quantities that are being calculated. For example, the distribution function represents the density of the model in 6D phase space, and the mass density profile can be recovered by integrating the distribution function over velocity space. In the case of an isotropic orbital structure, we find 
\begin{equation}
\rho(r) 
= 
4\!\sqrt2\,\pi \int_0^{\Psi(r)} f_{\text{iso}}(\calE) \sqrt{\Psi(r)-\calE}\,\txd\calE,
\label{dfcheckrho}
\end{equation}
and for an Osipkov-Merritt orbital structure we have \citep{1985AJ.....90.1027M}
\begin{equation}
\rho(r) 
= 
4\!\sqrt2\,\pi \left(1+\frac{r^2}{r_\txa^2}\right)^{-1}
\int_0^{\Psi(r)} f_{\text{om}}(Q) \sqrt{\Psi(r)-Q}\,\txd Q.
\label{dfcheckrhoQ}
\end{equation}
In a similar way, the velocity dispersion profiles can be found as the second-order moment of the distribution function, resulting in 
\begin{equation}
\rho(r)\,\sigma_{\text{iso}}^2(r) 
= 
\frac{8\!\sqrt2\,\pi}{3} 
\int_0^{\Psi(r)} f_{\text{iso}}(\calE) \left[\Psi(r)-\calE\right]^{3/2}\,\txd\calE,
\label{dfcheckrhosigma2}
\end{equation}
and 
\begin{multline}
\rho(r)\,\sigma_{r,{\text{om}}}^2(r) 
= 
\frac{8\!\sqrt2\,\pi}{3} \left(1+\frac{r^2}{r_\txa^2}\right)^{-1}
\\
\times\int_0^{\Psi(r)} f_{\text{om}}(Q) \left[\Psi(r)-Q\right]^{3/2}\,\txd Q.
\label{dfcheckrhosigmar2Q}
\end{multline}
The four Eqs.~(\ref{dfcheckrho})--(\ref{dfcheckrhosigmar2Q}) can easily be converted to integrations with respect to radius. For example, Eq.~(\ref{dfcheckrho}) can be transformed into
\begin{equation}
\rho(r) 
= 
4\!\sqrt2\,\pi G \int_r^\infty \frac{f_{\text{iso}}(\Psi(u))\,M(u) \sqrt{\Psi(r)-\Psi(u)}\,\txd u}{u^2}.
\end{equation}
We have implemented all of these inverse formulae in the {\tt{SpheCow}} code to be used as a way to validate the implementation of the different models. 

\begin{figure}
\includegraphics[width=0.45\textwidth]{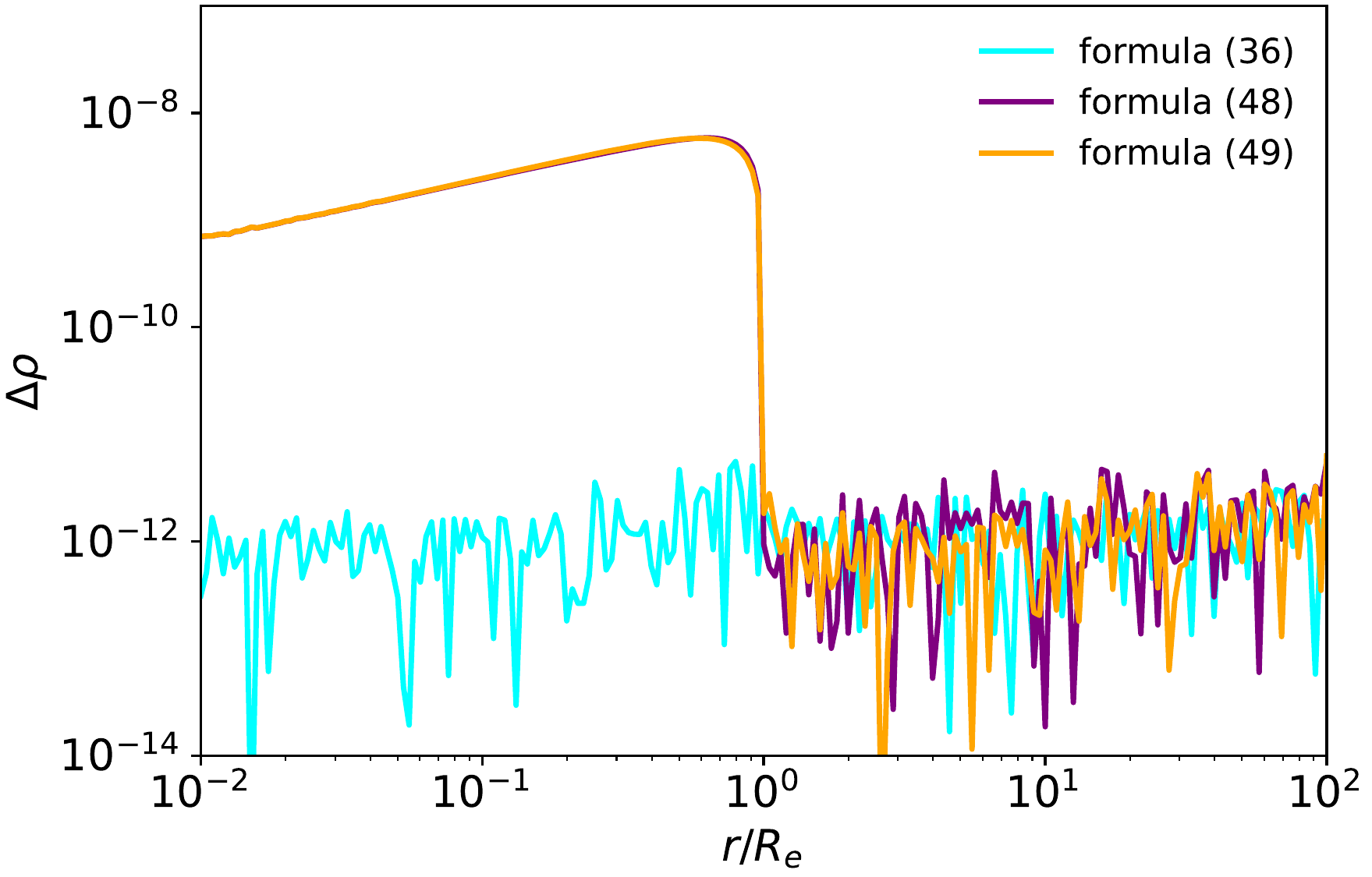}%
\caption{Absolute value of the relative error of the density of the Vaucouleurs model, as calculated by {\tt{SpheCow}}. The cyan line corresponds to the calculation using the standard deprojection equation (\ref{Itorho}), the purple line to the integration of the isotropic distribution function over velocity space (\ref{dfcheckrho}), and the orange line to the integration of the Osipkov-Merritt distribution function over velocity space (\ref{dfcheckrhoQ}). }
\label{Validation-DeVaucouleurs.fig}
\end{figure}

Moreover, they can also be used as a way to characterise the accuracy of the numerical integrations. In Fig.~{\ref{Validation-DeVaucouleurs.fig}} we show the absolute value of the relative error for the density of the de Vaucouleurs model. This model is implemented in the {\tt{SpheCow}} by means of its analytical surface density profile, and the cyan curves correspond to the density calculated via the standard deprojection formula (\ref{Itorho}). Comparing these values to the exact values, which can only be expressed exactly in terms of non-standard special functions \citep{2002A&A...383..384M, 2011A&A...525A.136B}, we see that the relative errors are of the order of $10^{-12}$, similar to what we obtained for the Plummer model in Fig.~{\ref{Validation-Plummer.fig}}. The purple lines in Fig.~{\ref{Validation-DeVaucouleurs.fig}} correspond to the density as calculated by integrating the isotropic distribution function over velocity space, that is using formula~(\ref{dfcheckrho}). The orange lines are similar, but now the Osipkov-Merritt distribution function is integrated over velocity according to Eq.~(\ref{dfcheckrhoQ}), with $r_\txa/R_\txe=2$. For both the isotropic and the Osipkov-Merritt cases, the relative error on the density for $r>R_{\text{e}}$ is of the same magnitude as the relative error of the density as calculated through the standard deprojection formula. This is remarkable, given that the integrands in the formulae~(\ref{dfcheckrho}) and (\ref{dfcheckrhoQ}) are a function of the distribution function, the mass profile and the potential, each of which are obtained via a numerical integration itself (the distribution function even via a complicated double integration). For $r<R_{\text{eff}}$, we see that there is a systematic increase of the relative error up to about $10^{-9}$, because the integration now involves a combination of two separate integrals.  

Finally, we can also test the implementation of the different models through the differential energy distribution. For a model with an isotropic orbital structure, $\calN(\calE)$ represents the distribution of mass as a function of the binding energy, and it should therefore satisfy the normalisation
\begin{equation}
\int_0^{\Psi_0} \calN_{\text{iso}}(\calE)\,\txd\calE 
= G\int_0^\infty \frac{f_{\text{iso}}(\Psi(u))\,g_{\text{iso}}(\Psi(u))\,M(u)\,\txd u}{u^2} = M.
\label{checkNE}
\end{equation}
Similarly, we have
\begin{equation}
\int_0^{\Psi_0} \calN_{\text{om}}(Q)\,\txd Q 
= G\int_0^\infty \frac{f_{\text{om}}(\Psi(r))\,g_{\text{om}}(\Psi(u))\,M(u)\,\txd u}{u^2} = M.
\label{checkNQ}
\end{equation}
All models currently in the {\tt{SpheCow}} code have successfully passed these tests (for different parameters for those models that depend on different parameters). We strongly recommend any user who adds a new model to the {\tt{SpheCow}} code, something we strongly encourage, to perform these basic tests as well.

\section{Sigmoid models}
\label{Sigmoid.sec}

\subsection{Density sigmoid models}

As an application and a demonstration of the power of the {\tt{SpheCow}} code, we develop a number of new families of spherical models. Many models, both for galaxies and dark matter haloes, are characterised by a density profile with a power-law behaviour at small and large radii, that is
\begin{subequations}
\label{powerlaw}
\begin{gather}
\rho(r) \propto r^{-\gamma} \qquad{\text{for }}r\ll r_\txb,
\\
\rho(r) \propto r^{-\beta} \qquad{\text{for }}r\gg r_\txb.
\end{gather}
\end{subequations}
with $r=r_\txb$ some break radius. An equivalent condition is that the logarithmic density slope $\gamma(r)$ behaves as 
\begin{subequations}
\label{gammaasy}
\begin{gather}
\gamma(r) \approx \gamma \qquad{\text{for }}r\ll r_\txb,
\\
\gamma(r) \approx \beta \qquad{\text{for }}r\gg r_\txb.
\end{gather}
\end{subequations}
Furthermore, it is desirable that the logarithmic density slope changes smoothly and monotonically between the asymptotic values $\gamma$ and $\beta$. Almost all of the models in the first part of the list in Table~{\ref{models.tab}} satisfy these criteria. A general and flexible model that was proposed with exactly these criteria is the Zhao or double power-law model, presented by \citet{1996MNRAS.278..488Z}. This model is used extensively to model dark matter haloes \citep[e.g.][]{2013MNRAS.433.2314H, 2014MNRAS.443.3712H, 2015ApJ...800...15H, 2014MNRAS.441.2986D, 2017A&A...605A..55A, 2017MNRAS.468.1005D, 2020MNRAS.491.4523F, 2020MNRAS.499.2912F, 2020ApJ...904...45H}. 

The family of Zhao models is, however, not the only possible family of models with the characteristic behaviour (\ref{gammaasy}). We can generate various new families of models with this behaviour using the concept of sigmoid functions. Sigmoid functions, generally denoted as $\calS(x)$, are monotonically increasing functions that map the entire real domain onto a finite range, which is usually taken to be the unit interval $[0,1]$. Moreover, sigmoid functions have a single inflection point, usually taken at $x=0$, such that $\calS(x)$ is convex for $x<0$ and concave for $x>0$.  Sigmoid functions play an important role in the theory of population dynamics in a wide variety of scientific disciplines \citep[e.g.][]{Richards1959, Tsoularis2002, banks2013growth, Fornalski2020}. Sigmoid function are also commonly used in artificial intelligence as activation functions in neural networks \citep{pmlr-v97-hayou19a, Hurbans2020}. Commonly used sigmoid functions include the logistic function, the arctangent function, and the Gompertz function; a number of possible options are listed in Table~{\ref{sigmoidfunctions.tab}}.

\begin{table}
\caption{Some of the most commonly used sigmoid functions. All these functions are normalised to have horizontal asymptotes at 0 and 1, and a single inflection point at $x=0$.} 
\label{sigmoidfunctions.tab}
\centering
\begin{tabular}{lc}
\hline\hline\\
name & $\calS(x)$ \\[1em]
\hline \\
logistic & $\displaystyle \frac{1}{1+\txe^{-x}}$ \\[1.2em]
hyperbolic tangent & $\displaystyle \frac{1+\tanh x}{2}$ \\[1.2em]
arctangent & $\displaystyle \frac12\left(1+\frac{2}{\pi}\arctan x\right)$ \\[1.2em]
error & $\displaystyle \frac{1+\erf x}{2}$ \\[1.2em]
Gompertz & $\displaystyle \exp\left(-\txe^{-x}\right)$ \\[1.2em]
Gudermannian & $\displaystyle \frac12\left[1+\frac{4}{\pi}\arctan(\tanh x) \right]$ \\[1.2em]
algebraic & $\displaystyle \frac12\left(1+\frac{x}{\sqrt{1+x^2}}\right)$ \\[1.2em]
Richards & $\displaystyle \frac{1}{(1+\nu\,\txe^{-x})^{1/\nu}} \qquad(\nu>0)$ \\[1.5em]
\hline\hline
\end{tabular}
\end{table}

Now consider any sigmoid function $\calS(x)$ and set
\begin{equation}
\gamma(r) = \gamma + (\beta-\gamma)\,\calS\left(\alpha \ln\left(\frac{r}{r_\txb}\right)\right).
\end{equation}
Interpreting this expression as the logarithmic density slope of a spherical model, we can find the density profile as
\begin{equation}
\rho(r) 
= 
\rho_\txc \left(\frac{r}{r_\txb}\right)^{-\gamma} 
\exp\left[-\frac{\beta-\gamma}{\alpha}\,\int_{-\infty}^{\alpha\ln(r/r_\txb)} \calS(x)\,\txd x\right],
\label{rhosigmoid}
\end{equation}
with $\rho_\txc$ an arbitrary constant. Due to the properties of the sigmoid function, we immediately know that the model with this density profile behaves exactly as expression~(\ref{powerlaw}). As we can apply the strategy above for any sigmoid function, we can easily generate various new classes of spherical models with this required asymptotic behaviour. As long as the integral in Eq.~(\ref{rhosigmoid}) can be evaluated analytically, this results in an analytical density profile. 

The obvious first option is to consider the logistic function, one of the most popular sigmoid functions, 
\begin{equation}
\calS(x) = \frac{1}{1+\txe^{-x}}.
\label{logistic}
\end{equation}
This logistic function was originally introduced to describe the evolution of the Belgian population \citep{Verhulst1845}, and is sometimes referred to in the literature as {\em{the}} sigmoid function. Combining Eqs.~(\ref{rhosigmoid}) and (\ref{logistic}) leads to the following expressions for the logarithmic density slope,
\begin{equation}
\gamma(r) = \frac{\gamma+\beta\,(r/r_\txb)^\alpha}{1+(r/r_\txb)^\alpha},
\end{equation}
and the density profile,
\begin{equation}
\label{rho-Zhao}
\rho(r) = \rho_\txc \left(\frac{r}{r_\txb}\right)^{-\gamma} 
\left[1+\left(\frac{r}{r_\txb}\right)^\alpha\right]^{-\frac{\beta-\gamma}{\alpha}}.
\end{equation}
This is nothing but the general expression of the density profile of the Zhao model. 

\begin{figure*}
\includegraphics[width=0.95\textwidth]{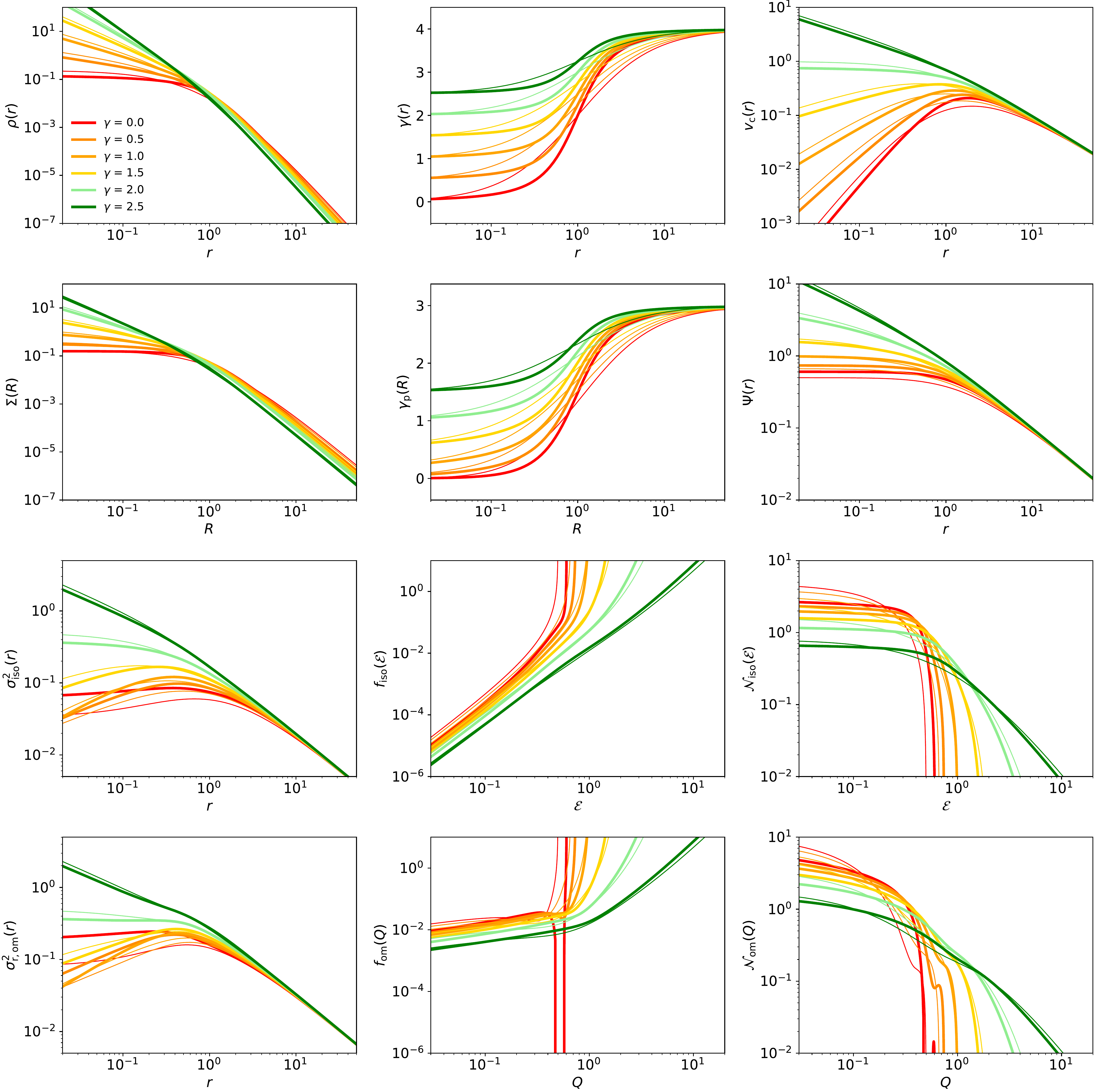}%
\caption{Compilation of photometric and dynamical properties for two families of sigmoid models. The thick lines correspond to the new family of algebraic sigmoid models defined by the density profile (\ref{SigmoidDensity-rho}). All models have $\alpha=1$ and $\beta=4$, and the different colours correspond to different values of the inner density slope $\gamma$, as indicated in the top left panel. The thin lines correspond to Zhao models with exactly the same parameters.}
\label{SigmoidDensity.fig}
\end{figure*}

\subsection{Algebraic sigmoid density models}

As an alternative to the logistic function, consider the algebraic sigmoid function,
\begin{equation}
\label{algesigmoid}
\calS(x) = \frac12\left(1+\frac{x}{\sqrt{1+x^2}}\right). 
\end{equation}
This choice leads to a new family of models with logarithmic density slope 
\begin{equation}
\gamma(r) = \frac{\beta+\gamma}{2} + \frac{\beta-\gamma}{2}\frac{\alpha\ln (r/r_\txb)}{\sqrt{1+\alpha^2\ln^2 (r/r_\txb)}},
\label{SigmoidDensity-gamma}
\end{equation}
and density profile
\begin{equation}
\rho(r) = \rho_\txc \left(\frac{r}{r_\txb}\right)^{-\frac{\beta+\gamma}{2}} 
\exp\left[-\frac{\beta-\gamma}{2\alpha}\sqrt{1+\alpha^2\ln^2\left(\frac{r}{r_\txb}\right)}\right].
\label{SigmoidDensity-rho}
\end{equation}
We have added this family of models to the {\tt{SpheCow}} code as the {\tt{SigmoidDensityModel}} class. The only member functions we had to implement were the density and its first and second derivatives. All other photometric and dynamical properties of this family of models can be evaluated without any further ado. 

In Fig.~{\ref{SigmoidDensity.fig}} we compare the dynamical properties of a subset of the algebraic sigmoid models defined by the density profile~(\ref{SigmoidDensity-rho}) and the Zhao models. The different panels represent a wide variety of profiles, all of which are automatically calculated by {\tt{SpheCow}}. The thick lines in this figure represent the models from the new family of algebraic sigmoid models. For all models, we have fixed $M_{\text{tot}} = r_\txb =1$, $\alpha=1$, and $\beta=4$, and we have considered different values of the inner slope $\gamma$. The thin lines correspond to Zhao models with exactly the same parameters. Actually, the subfamily of Zhao models with $(\alpha, \beta) = (1,4)$ is the well-known family of $\gamma$-models \citep{1993MNRAS.265..250D, 1994AJ....107..634T}.

It is clear that corresponding models from the two families have the same asymptotic behaviour in all quantities, as expected as they have the same asymptotic density profile by construction. The properties of the models of both families display a significant variety at small radii, depending on the parameter $\gamma$ that characterises the inner density slope. For $\gamma=0$ the models have a finite central density, surface density, potential, and isotropic velocity dispersion. For $0<\gamma<1$ both families of models show a shallow density cusp, but still a finite central surface density and finite potential well. The isotropic dispersion profile now tends to zero at small radii. Models with $1<\gamma<2$ show a similar behaviour, except for their surface density profile which diverges in the centre. Models with $\gamma>2$ have a strong density and surface density cusp, an infinitely deep potential well, and diverging circular velocity and isotropic dispersion profiles. 

While the asymptotic behaviour of the corresponding Zhao and algebraic sigmoid models is the same, there are some minor differences in the details. For a fixed set of parameters, the algebraic sigmoid models are characterised by a steeper transition between the inner and outer profiles, as shown in the top left and top middle panels. This difference propagates to all other properties as well. Algebraic sigmoid models with small values of $\gamma$ have a deeper potential well than the corresponding $\gamma$-models, and the opposite is true for large values of $\gamma$. Looking at the isotropic distribution function and differential energy distribution, the transition between the low and high binding energy parts is much sharper for the algebraic sigmoid models. As a result, the Zhao models have a larger number of stars with low binding energies, whereas the algebraic sigmoid models have more stars around the knee of the differential energy distribution. The specific choice of the sigmoid function does matter for the details of the dynamical structure. 

Focusing on the panels on the third row of Fig.~{\ref{SigmoidDensity.fig}} we see that all algebraic sigmoid models shown have an isotropic distribution function that is a positive and monotonically increasing function of binding energy. This implies that these models are self-consistent and physical, and stable against radial and non-radial perturbations \citep{1962spss.book.....A, 1971PhRvL..26..725D, 2008gady.book.....B}. We do note, however, that this conclusion cannot be generalised for all members of the family of algebraic sigmoid models, in a similar way as for the family of Zhao models \citep{2021MNRAS.503.2955B}. If the parameter $\alpha$ that determines the sharpness of the transition between the inner and outer profiles grows larger, the distribution function will first develop a kink around $\calE=\Psi(r_\txb)$. When $\alpha$ increases even more, the distribution function will become negative, making the models with an isotropic orbital structure unphysical. In the limit $\alpha\to\infty$, the models reduce to the broken power-law (BPL) models discussed by \citet{2020ApJ...892...62D} and \citet{2021MNRAS.503.2955B}. These models are characterised by the simple density profile
\begin{equation}
\rho(r) = \begin{cases}
\;\rho_\txc \left(\dfrac{r}{r_\txb}\right)^{-\gamma} & \qquad{\text{if }}r\leqslant r_\txb, \\[1em]
\;\rho_\txc \left(\dfrac{r}{r_\txb}\right)^{-\beta} & \qquad{\text{if }}r\geqslant r_\txb.
\end{cases}
\end{equation}
This density profile is indeed what we obtain if we take the limit $\alpha\to\infty$ in expression~(\ref{SigmoidDensity-rho}). The BPL models can also be seen as sigmoid models with the Heaviside step function 
\begin{equation}
\calS(x) = \Theta(x) \equiv \begin{cases}
\;0 & \qquad{\text{if }}x<0, \\
\;\tfrac12 & \qquad{\text{if }}x=0, \\
\;1 & \qquad{\text{if }}x>0,
\end{cases}
\end{equation}
as their defining (degenerate) sigmoid function. 

Finally, it is interesting to compare the panels on the last two rows of Fig.~{\ref{SigmoidDensity.fig}}. These rows show similar quantities, but the third row corresponds to an isotropic orbital structure, while the bottom row shows the same properties for an Osipkov-Merritt orbital structure with $r_\txa=0.6$. At large $\calE$ or $Q$, the isotropic and Osipkov-Merritt distribution functions for any given set of parameters show a similar behaviour. This is not so surprising, as large binding energies correspond to small radii, and the Osipkov-Merritt models are isotropic in the central regions. The behaviour at small $\calE$ or $Q$, corresponding to large radii, is very different, with a different slope for both the distribution function and the differential energy distribution. The most interesting difference, however, relates to the behaviour at intermediate binding energies. The anisotropy generates a kink in the distribution function and the differential energy distribution, especially for models with small values of $\gamma$. For the algebraic sigmoid model with $\gamma=0$, this kink is so strong that the distribution function and the differential energy distribution become negative: this model cannot be supported by an Osipkov-Merritt orbital structure with $r_\txa=0.6$. We note that the corresponding $\gamma$-model, that is the one with $\gamma=0$ and $r_\txa=0.6$, is also affected by the anisotropy, but the distribution function remains positive over the entire domain. This is in agreement with the results of \citet{1995MNRAS.276.1131C}, who found that the $\gamma=0$ model can be supported by an Osipkov-Merritt orbital structure as long as $r_\txa>0.44$.

\subsection{Surface density sigmoid models}

One of the useful characteristics of the {\tt{SpheCow}} code is that it is essentially equally simple to generate models by means of a density profile or a surface density profile. Actually, for every model based on a certain density profile, we can immediately generate a counterpart with a surface brightness profile with a similar functional form, and vice versa. One well-known example of such a couple of twin models consists of the Einasto and S\'ersic models. We note that the families of Einasto and S\'ersic models have no connection to each other apart from a functional similarity between the density profile of the former and the surface density profile of the latter. In particular, Einasto models do not have a S\'ersic surface density profile, and vice versa \citep{2010MNRAS.405..340D, 2012A&A...540A..70R}.

In the same sense, it is also straightforward to generate new families of models in which the logarithmic slope of the surface density profile has the shape of a sigmoid function. For any sigmoid function $\calS(x)$, we set
\begin{equation}
\gamma'(R) = \gamma + (\beta-\gamma)\,\calS\left(\alpha \ln\left(\frac{R}{R_\txb}\right)\right),
\end{equation}
resulting in a surface density profile
\begin{equation}
\Sigma(R) 
= 
\Sigma_\txc\left(\frac{R}{R_\txb}\right)^{-\gamma} 
\exp\left[-\frac{\beta-\gamma}{\alpha}\,\int_{-\infty}^{\alpha\ln (R/R_\txb)} \calS(x)\,\txd x\right],
\end{equation}
with a power-law behaviour at both small and large projected radii. If we take the logistic function as the standard sigmoid function, this results in the surface density profile,
\begin{equation}
\label{Sigma-Nuker}
\Sigma(R) = \Sigma_\txc \left(\frac{R}{R_\txb}\right)^{-\gamma} 
\left[1+\left(\frac{R}{R_\txb}\right)^\alpha\right]^{-\frac{\beta-\gamma}{\alpha}}.
\end{equation}
This surface density profile defines the family of Nuker models, frequently used to model the nuclear regions of galaxies \citep[e.g.][]{1995AJ....110.2622L, 2005AJ....129.2138L, 2007ApJ...664..226L, 1996AJ....111.1889B, 2000ApJS..128...85Q, 2001AJ....121.2431R, 2003AJ....125..478L, 2005A&A...439..487D, 2019A&A...622A..78D}. Similar to the Einasto--S\'ersic couple of models, the Zhao and Nuker models have no connection to each other except the similarity between the density profile (\ref{rho-Zhao}) of the former and the surface density profile (\ref{Sigma-Nuker}) of the latter \citep[see also][]{1997MNRAS.287..525Z}.

Any other sigmoid function is possible as well, and will lead to a different family of models with slightly different photometric and dynamical properties. If we use the algebraic sigmoid function (\ref{algesigmoid}), we find
\begin{equation}
\Sigma(R) = \Sigma_\txc \left(\frac{R}{R_\txb}\right)^{-\frac{\beta+\gamma}{2}} 
\exp\left[-\frac{\beta-\gamma}{2\alpha}\sqrt{1+\alpha^2\ln^2\left(\frac{R}{R_\txb}\right)}\right].
\label{SigmoidSurfaceDensity-Sigma}
\end{equation}
We have implemented the family of models defined by this surface density profile as the {\tt{SigmoidSurfaceDensityModel}} class in the {\tt{SpheCow}} code. This required just a few minor adaptations from the implementation of the {\tt{SigmoidDensityModel}} class: the main adaptation is the required implementation of the third derivative of the surface density profile. 

We do note provide an exhaustive overview or description of the dynamical properties of this new class of spherical models. It is not surprising that this general family of models shares many properties with the family of Nuker models, which has the same asymptotic behaviour. Just as in the previous subsection, there are minor differences in the surface density profile between corresponding models from both families, which propagate to all other properties. For a more detailed discussion of the Nuker models, we refer to \citet{2020A&A...634A.109B}, where we used a combination of analytical tools and a preliminary version of {\tt{SpheCow}} to investigate the dynamical structure. 

\section{Discussion and summary}
\label{Summary.sec}

The goal of this work was to develop and describe a new software tool that can be used to set up a spherical dynamical model with either an arbitrary analytical density profile or an arbitrary analytical surface density profile as starting point. The ambition was that this code should be capable of automatically calculating the basic intrinsic and on-sky properties for any such model, as well as dynamical properties assuming either an isotropic of an Osipkov-Merritt type anisotropic orbital structure. With such a code, it should be easy to replace the standard but not necessarily realistic analytical models for galaxies or dark matter haloes that are often used as starting points for theoretical or numerical investigations by more optimal alternatives.

The result of this work is {\tt{SpheCow}}, a light-weight and flexible software code that has exactly the characteristics as described above. We have gathered all the relevant equations, and through some integral order rearrangements, derived some new expressions that make these calculations more simple. We have described our numerical integration strategy, which is based on a straightforward but very efficient Gauss-Legendre scheme. We have extensively validated {\tt{SpheCow}} using a combination of comparisons to analytical calculations, and the calculation of inverse formulae. The code is publicly available as a set of C++ routines and as a Python module.

{\tt{SpheCow}} contains readily usable implementations for many standard models, including the Plummer, Hernquist, NFW, Einasto, S\'ersic, and Nuker models. The main strength of the code is, however, that it is  designed to be easily extendable, in the sense that new models, defined by either an analytical density profile or an analytical surface density profile, can be added in a straightforward way. We demonstrate this by adding two new families of models: one with its logarithmic density slope and the other with its logarithmic surface density slope described by an algebraic sigmoid function. We do not investigate the properties of these models in full detail, but we show how straightforward it is to implement new models and how easy it is to subsequently explore their dynamical structure.

Apart from adding new models, the {\tt{SpheCow}} code can be expanded in several other ways. One limitation of the current implementation is that only two orbital structures, isotropic and Osipkov-Merritt, are available. Other options are possible and probably desirable. In particular, neither the isotropic nor the Osipkov-Merritt case provides an accurate representation of the orbital structure of dark matter haloes as obtained from simulations \citep{1996MNRAS.281..716C, 2000ApJ...539..561C, 2005MNRAS.363..705M, 2006NewA...11..333H}. Several other representations for the anisotropy profile have been proposed in the literature, which provide a more suitable fit to the results of N-body simulations \citep[e.g.][]{1997ApJ...485L..13C, 2004MNRAS.352..535D, 2005MNRAS.363..705M, 2007A&A...476L...1T, 2010MNRAS.401.2433M}. Unfortunately, in the general case, the calculation of the distribution function for such general anisotropy profiles is cumbersome and ill-conditioned \citep{1986PhR...133..217D}. There are, however, a number of special cases for which the inversion can be performed in a similar way as for the isotropic and Osipkov-Merritt models. The most obvious case are models with a constant anisotropy \citep{1986PhR...133..217D, 2006ApJ...642..752A}, and in particular the special cases of models populated only with radial orbits \citep{1984ApJ...286...27R}, or models with a constant radial anisotropy $\beta=\tfrac12$ \citep{2006PhRvD..73b3524E}. Other, more complex orbital structures for which the inversion can be performed (numerically) include the models by \citet{1991MNRAS.253..414C} and \citet{1995MNRAS.275.1017C}. We plan an implementation in {\tt{SpheCow}} of dynamical models with some of these orbital structures as future work, and invite third parties to contribute to this effort.

This is not the only possible way in which {\tt{SpheCow}} can easily be expanded. All models implemented at this stage are self-consistent self-gravitating models, that is, the density and gravitational potential are linked by Poisson's equation. It is also possible to drop this restriction and to include models consisting of two or more components. Such models can be useful, for example, to study the effect of a central supermassive black hole and/or a dark matter halo on the stellar dynamical structure of a galaxy \citep[e.g.][]{1994AJ....107..634T, 1996ApJ...471...68C, 1996MNRAS.282....1C, 2019MNRAS.490.2656C, 2004MNRAS.351...18B, 2005A&A...432..411B, 2018MNRAS.473.5476C}. 

We hope that this paper will inspire colleagues to use {\tt{SpheCow}}, both to investigate the properties of existing models in detail and to explore models that have so far received less attention than they deserve.

\begin{acknowledgements}
M.B.\ and B.V.\ acknowledge financial support from the Belgian Science Policy Office (BELSPO) through the PRODEX project “SPICA-SKIRT: A far-infrared photometry and polarimetry simulation toolbox in preparation of the SPICA mission” (C4000128500). The authors thank the anonymous referee for a prompt and constructive referee report.
\end{acknowledgements}

\bibliography{SpheCow}

\end{document}